\documentclass{aa}
\usepackage{graphicx}

\usepackage{txfonts}

\newcommand{\Te}{$T_{\rm{e}}$}

\newcommand{\Ha}{H$\alpha$}
\newcommand{\Hb}{\ifmmode {\rm H}\beta \else H$\beta$\fi}

\newcommand{\Nii}{[N\,{\sc ii}]\,$\lambda$6584}

\newcommand{\nii}{[N\,{\sc ii}]}

\newcommand{\Oii}{[O\,{\sc ii}]\,$\lambda$3727}

\newcommand{\Oiii}{[O\,{\sc iii}]\,$\lambda$5007}
\newcommand{\Oiiit}{[O\,{\sc iii}]\,$\lambda$4363}

\newcommand{\Siii}{[S\,{\sc iii}]\,$\lambda$9069}

\newcommand{\Ariii}{[Ar\,{\sc iii}]\,$\lambda$7135}
\newcommand{\ariii}{[Ar\,{\sc iii}]}

\newcommand{\rOiii}{[O\,{\sc iii}]\,$\lambda$4363/5007}

\begin{document}

     \title{[Ar\,{\sc iii}]/[O\,{\sc iii}] and [S\,{\sc iii}]/[O\,{\sc iii}]: well-behaved oxygen abundance indicators for HII regions and star forming galaxies}

     \author{Gra\.zyna Stasi\'nska\inst{1}}

     \offprints{G. Stasi\'nska, \email{grazyna.stasinska@obspm.fr}}

     \institute{LUTH, Observatoire de Paris-Meudon, 5 Place Jules Jansen, F-92195 Meudon, France}
          \date{Received ???; accepted ???}
     \titlerunning{[Ar\,{\sc iii}]/[O\,{\sc iii}] and [S\,{\sc iii}]/[O\,{\sc iii}] }
\authorrunning{ }


          \abstract{We propose two statistical methods to derive oxygen abundances in HII regions and star forming galaxies and calibrate them with a sample of several hundred giant HII regions in spiral and blue compact galaxies as well as of galaxies from the Sloan Digital Sky Survey. We  show the advantages of our new abundance indicators over previous ones.

}

         \maketitle

\section{Introduction}

Emission lines due to photoionization of nebulae by massive stars are the most powerful indicators of the chemical evolution of galaxies in both the near and the intermediate redshift universe. The wealth of new data coming from large/deep galaxy spectroscopic surveys and from observations with integral field units on nearby galaxies should allow a detailed inventory of the chemical composition of star forming galaxies.

Abundance determinations in HII regions are relatively straightforward if the electron temperature, \Te, can be measured directly from the observations (although strong biases  may affect  abundance derivations in high metallicity objects, even if lines used for \Te\  measurements  are observed, see   Stasi\'nska 2005). In many cases, however, \Te\ diagnostic lines are too faint to be detected. Pagel et al. (1979) and Alloin et al. (1979) pioneered methods that allow one to estimate the metallicity\footnote {Throughout the paper the word ``metallicity'' is used with the meaning of  ``oxygen abundance'' } using  strong lines only. These methods assume that all  HII regions  are essentially characterized by their metallicity, other parameters such as the hardness of the ionizing radiation and the ionization parameter being actually linked to metallicity. Empirically, this proves true, at least statistically, as has been shown by many theoretical and observational studies. 
The biggest problem is the translation of an observed line ratio into metallicity. We will not review here the large amount of literature devoted to this subject. References can be found in P\'erez-Montero \& D\'{i}az (2005), Pettini \& Pagel (2004, hereafter PP04) and Stasi\'nska (2004). There are two important issues. One is the choice of the metallicity indicator, the other is the proper  abundance calibration. Here, we mostly discuss the first of the two, as we believe that time is not ripe yet to discuss the second: one really must understand the physics of metal-rich HII regions to be confident in any calibration at high metallicity. This requires the systematic use of mid-infrared data to efficiently constrain the modelling of metal-rich HII regions.

\begin{figure*}\includegraphics[width=12cm]{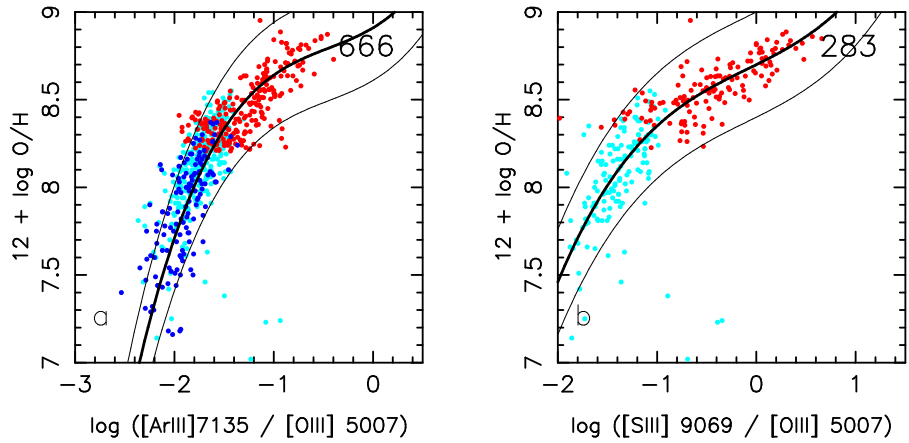} 
\caption{{\bf a} 12+log O/H as a function of log $Ar _3O_3$.   Red:   HII regions from the spiral galaxies sample; blue:  HeBCD sample;   cyan: SDSS DR3 sample.  The thick curve represents our calibration defined by Eq. (1). The two thin curves are vertically displaced from the thick one by $\pm$0.3 dex.  {\bf b}    Same for 12+log O/H as a function of log $S _3O_3$. The thick curve represents our calibration  defined by  Eq. (2). }\label{fig1}\end{figure*}

The most widely used indexes are the following:  $O_{23}$, i.e. ({[O\,{\sc iii}]\,$\lambda$4959,5007}+\Oii)/\Hb , introduced by Pagel et al. (1979), then refined by Mc Gaugh (1991) and Pilyugin (2001a);  $O_3N_2$, i.e.  \Oiii/\Nii\ introduced by Alloin et al. 1979 and reconsidered by PP04; $ N_2$, i.e. \Nii/\Ha\   first proposed by Storchi-Bergmann et al. (1994) and Van Zee et al. (1998), and  last recalibrated by PP04. Another index that is becoming popular, although it requires observations in the far red, is  $S_{23}$ , i.e. ({[S\,{\sc iii}]\,$\lambda$9069}+{[S\,{\sc ii}]\,$\lambda$6716,6731})/\Ha\  (V\'{i}lchez \& Esteban 1996, P\'erez-Montero \& D\'{i}az 2005). 
The main problem of the $O_{23}$ index is that it is double valued. Therefore another index is needed to decide whether one has to chose the low-metallicity or the high-metallicity solution. This is not practical if dealing with  data sets  where both high and low metallicities are expected to be present. The $S_{23}$ index is better since it is monotonic up to about solar metallicity, but a bend at higher metallicities is expected. The $O_3N_2$ and $ N_2$ indexes, on the other hand, have the virtue of being single valued. However, the reasons for this are not purely ``physical'' but are partly due to ``astrophysical'' causes:  it is an observed fact that,  as metallicity increases, both the hardness of the ionizing radiation and the ionization parameter of HII regions tend to decrease, and that the N/O ratio increases on average (at least at high metallicities).  The important role of the ``astrophysical factors'' ,  especially the variation of the N/O ratio  should invite us to some caution when relying on them to derive metallicities of objects that might have a different nature and  history than the objects used to calibrate the relation. Such might be the case of high redshift galaxies as compared to local dwarf galaxies.

Another problem that affects both the  $O_3N_2$ and $ N_2$ indexes is that they use a ``low excitation line'', \Nii\, which may arise not only in \textit{ bona fide} HII regions, but also in the diffuse ionized medium. It is estimated that the diffuse ionized medium  contributes  up to  one half of the total \Ha\ emission in galaxies (Thilker et al. 2002). Therefore,  calibrations of the  $O_3N_2$ and $ N_2$ indexes based on giant HII regions might not be relevant for the interpretation of integrated spectra of galaxies.

Ideally, one would require from a metallicity index to:  \emph {i)} be single valued with respect to metallicity,  \emph {ii)} have a behaviour dominated by a well understood ``physical'' reason,  \emph {iii)}  be unaffected by the presence of   diffuse ionized gas, and \emph {iv)}   be independent of  chemical evolution. We propose two indexes: $A _3O_3$ , i.e. \Ariii/\Oiii\ and  $S _3O_3$, i.e. \Siii/\Oiii\, that fulfill these requirements. In Sect.  2 we describe the data base sustaining this work, in Sect. 3, we present our new metallicity calibrators, and in Sect. 4, we discuss their merits  with respect to the usual ones.

\begin{figure}\includegraphics[width=5.5cm]{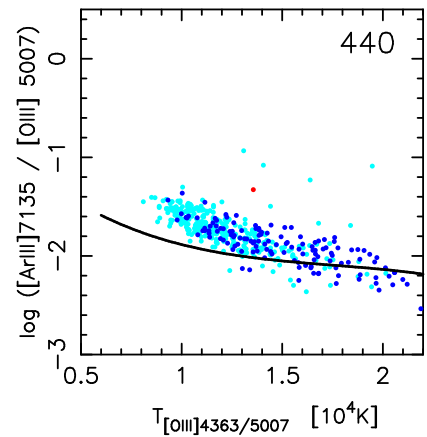}
\caption{ log $A_3O_3$ versus the electron temperature measured from the \rOiii\ ratio.   Blue:  HeBCD sample;   cyan: SDSS DR3 sample. The curve shows the logarithm of the \Ariii/\Oiii\   \emph {emissivity} ratio at the temperature given by the abscissa (the curve has been shifted for easier comparison with observational points). }\label{fig2}\end{figure}

\section{Data samples}

We have built a large -- though not exhaustive -- data base of published reddening-corrected  line intensities. Data for HII regions in spiral galaxies  have been taken from  Garnett \& Kennicutt (1994),  Garnett et al. (1997), van Zee et al. (1998), Bresolin et al. (1999, 2004,  2005),  and Kennicutt et al. (2003). We have also used the  sample of blue compact galaxies, observed by Izotov and coworkers for the quest of the pregalactic helium abundance, and referred to as the HeBCD sample in Izotov et al. (2006). The spectra in this sample have very good signal-to-noise ratios. Unfortunately, the wavelength range  does not reach the {[S\,{\sc iii}]\,$\lambda$9069} line. Finally, we have  used  data from the SDSS DR3 sample, i.e. data from the 3rd data release of the Sloan Digital Sky Survey (Abazajian et al. 2005) that have been selected by Izotov et al. (2006) and are listed in their Table 1.  

\begin{figure*}\includegraphics[width=12cm]{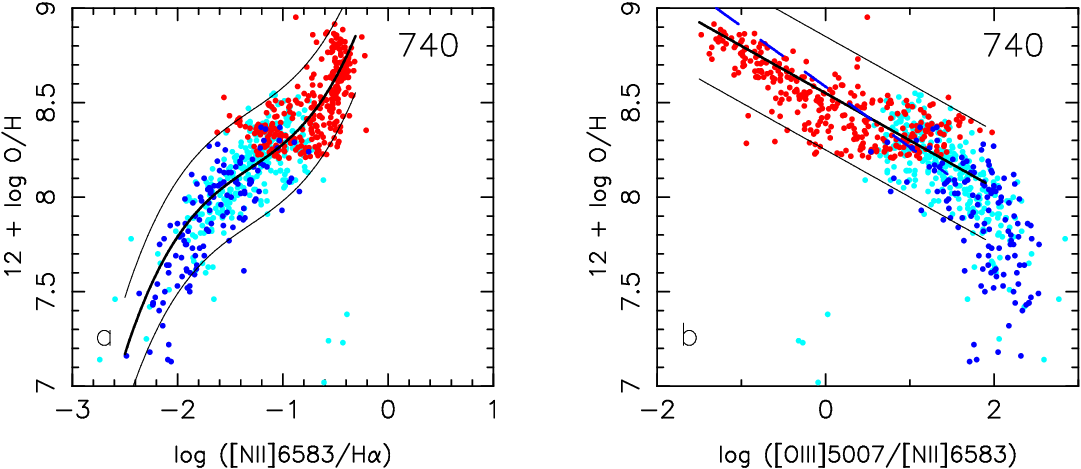} 
\caption{{\bf a} 12+log O/H vs. log $N_2$ for our data sample. The thick curve represents  Eq. (3).  {\bf b}  12+log O/H vs. log $O_3N_2$. The dashed line shows the PP04 calibration.The thick line is our proposed calibration (Eq. 4). The thin curves are vertically displaced from the thick one by $\pm$0.3 dex.  
}\label{fig1}\end{figure*}

For both the HeBCD and  SDSS DR3 samples, the oxygen abundances were computed using the temperature derived from \rOiii\ and following the prescriptions given in Izotov et al. (2006). The formal {\bf average} uncertainty in the computed 12+log O/H values is typically 0.09 dex for the SDSS DR3 sample and 0.03 for the HeBCD sample. Concerning  HII regions in spirals, many of which are located in the inner parts of the galaxies, only 22 objects out of over 300 have the intensity of \Oiiit\ Êmeasured. In some cases, other weak auroral lines have been measured. We have however decided not to adopt any \Te-based method, but rather to use the strong line method of Pilyugin (2001a) for high metallicities, which uses  a combination of $O_{23}$ and \Oiii/\Oii\ line ratios (the $P$-method). This method is based on a thorough consideration of many factors, it has been widely discussed (Pilyugin 2001b, 2003, Cedr\'es et al. 2004) and it is largely used. It is not clear at present whether more recent calibrations of the method  (Pilyugin \& Thuan 2005, Pilyugin et al. 2006) are more accurate, because of the problem of abundance bias at high metallicity  discussed in Stasi\'nska (2005). Since, in any event, abundances at high metallicities will have to be  reconsidered once systematic measurements of mid-infrared lines become available, we simply took the most reasonable of the widely used methods, i.e. the calibration by Pilyugin (2001a). We discarded the objects for which the method gives 12 + log O/H $\le$ 8.2, which is the limit of validity defined by Pilyugin for the high metallicity range. Pilyugin considers his method to be accurate within 0.1dex. However, the accuracy is certainly not as good near the bend in the $O_{23}$ versus 12+log O/H relation.

\section{ $Ar _3O_3$  and  $S _3O_3 $ as  metallicity calibrators}

We show in Fig. 1a the values of  12 + log O/H as a function of log $A_3O_3$. Red dots correspond to HII regions from our spiral galaxies sample, blue ones correspond to the HeBCD sample,  and cyan ones to the SDSS DR3 sample. In total, there are 666 objects in the plot. Apart from a few outliers in the SDSS DR3 sample (for which the oxygen abundance has likely been grossly underestimated, and which  deserve detailed study), all the objects from the three samples merge into one monotonic sequence, as expected.  The Spearman rank correlation coefficient, $R_S$, is  0.81. The observed trend of  $Ar _3O_3 $ is for a large part due to the decrease of \Te\ as metallicity increases,  which translates into an increase of  the  \Ariii\ and \Oiii\  emissivity ratio (see Fig. 2). The  fact that the hardness of the radiation field and ionization parameter decrease with increasing metallicity adds to the observed relation between   12 + log O/H and log $Ar _3O_3$, but does not play a dominant role.  Also, as expected from stellar nucleosynthesis and confirmed by observations of metal-poor HII regions (Izotov et al. 2006), Ar and O are produced by the same stars, so, unlike    $O_3N_2$ or $ N_2$ , the $Ar _3O_3 $ index is not affected by chemical evolution effects. Note that from Fig. 2 one can infer that  the largest observed values of $Ar _3O_3$ seen in Fig. 1a must correspond either to a rapid change in the hardness of the radiation field and/or ionization parameter at the high metallicity end, or to a strong temperature gradient inside the HII regions.

From  Fig. 1a, we can define a new metallicity calibration. The observed distribution of points in this figure is   well fitted by the following expression: 
\begin{equation}
12 + {\rm log~O/H}  =  8.91 + 0.34 x + 0.27 x^2 + 0.20 x^3
\end{equation}
where $x$=  log $Ar _3O_3$. The heavy line in Fig. 1a represents this curve. The standard deviation between log O/H ($O_{23}$) and log O/H ($A_3O_3$) using this curve is 0.23 dex. Of course, Eq. (1) may need revision when the physics of metal-rich HII regions becomes better understood.

Fig. 1b shows the values of 12 + log O/H as a function of log $S _3O_3$. Here, the number of data points is much lower (283). Especially, there are no data for the HeBCD sample. But the Spearman rank correlation coeffcient is still very high ($R_S$=0.84). The trend shown by the observational points is similar to the one seen if Fig. 1a, with a smaller slope, due to the larger dependence of the ratio of \Siii\ and \Oiii\ emissivities on electron temperature. The observed distribution of points can be fitted by:

\begin{equation}
12 + {\rm log~O/H} =  8.70  + 0.28x + 0. 03 x^2 + 0.1 x^3
\end{equation}
where $x$=  log $S _3O_3$. This is the heavy line in Fig. 1b.  The standard deviation between log O/H ($O_{23}$) and log O/H ($S_3O_3$) using this curve is 0.25 dex. 

Both Eqs. (1) and (2) where found  by eye estimate. We checked that, at least with our  sample, the derivations of metallicity using both methods are consistent: there is no specific trend in the differences between both values of log O/H and the standard deviation of the differences is   0.09 dex (see Fig. 4a).

\begin{figure*}\includegraphics[width=12cm]{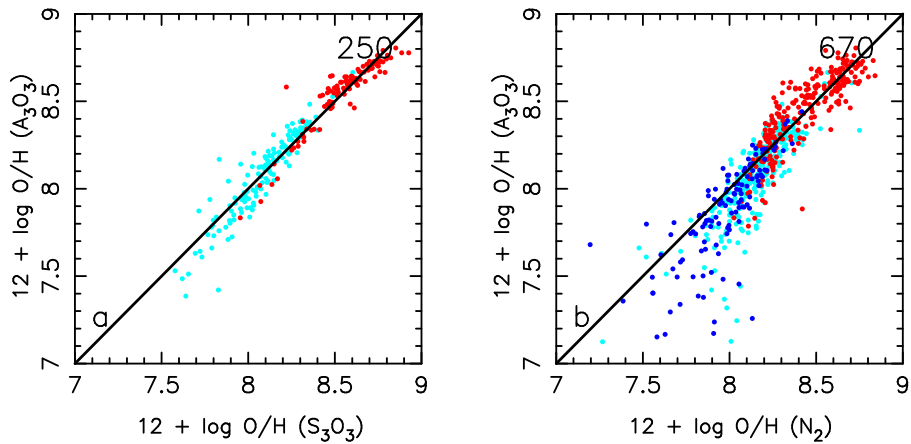}\caption{{\bf a} 12+log O/H derived from $Ar _3O_3 $ (Eq. 1) vs.   12+log O/H  derived from $S _3O_3 $ (Eq. 2).  {\bf b}  12+log O/H derived from $Ar _3O_3 $ (Eq. 1) vs.   12+log O/H  derived from $N _2 $ using the formula of PP04 (Eq. 3). The black lines show the 1:1 relations. }\label{fig1}\end{figure*}

\section{Discussion}

We now discuss the relative merits of our $Ar _3O_3 $ and $S _3O_3 $ indexes with respect to $O_3N_2$ and $ N_2$. Rather than comparing our Figs. 1a and 1b with Figs. 1 and 2 of PP04 (who use a different data base and different derivations of O/H), we plot in Figs. 3a and 3b the values of 12 + log O/H vs log $ N_2$ and  log $O_3N_2$, respectively, using our  samples. Note that our data points are more numerous (740 points instead of 137) and better distributed in the entire range of metallicities than the data points in PP04.  The heavy line in Fig. 3a represents the calibration of the $ N_2$  index as given by PP04: 
\begin{equation}
12 +{ \rm log~O/H} =  9.37  + 2.03 x +1.26 x^2 + 0.32  x^3
\end{equation}
where $x$=  log $N_2$. As can be seen, their calibration fits our data well. On the other hand, their expression relating 12 + log O/H with the $O_3N_2$ index, shown with the dashed blue line in Fig.3b, is a very bad fit to our data. This stems from the fact that their relation is heavily weighted by only 4 objects in which the oxygen abundance has been determined from photoionization modelling and was found to be high. Unfortunately, abundances derived from models are very uncertain if constraints are insufficient. It is also very puzzling that those 4 points are quite separated from the rest of the data in the (12 + log O/H vs. $O_3N_2$)  diagram of PP04, while they are not in their (12 + log O/H vs. $N_2$) diagram.  Anyway, our Fig. 3b clearly confirms their conclusion that  $O_3N_2$ alone is of no use for low metallicity objects. It can  be used for high metallicity objects, though, rather than the calibration proposed by PP04 (their Eq. 3)\footnote{Note that PP04 use   (\Oiii\Hb)/(\Nii/\Ha)  while we defined $O_3N_2$ as \Oiii/\Nii.}  we prefer:
\begin{equation}
12 +{ \rm log~O/H} =  8.55  - 0.25 x 
\end{equation}
where $x$=  log $O_3N_2$. However, comparison of  Fig. 3b with Figs. 1a,b shows that   there is no advantage in using $O_3N_2$. Given the other shortcomings of  $O_3N_2$ (see Sect. 1),  it is preferable to use our new indexes, especially  $Ar _3O_3 $, which does not require a large wavelength range. Note that \ariii\  was not considered as a ``strong line'' in previous studies but its intensity is comparable to that of \nii\ at low metallicities. The 
 $Ar _3O_3$ index is particularly well-suited to study the metallicities of the galaxies in the SDSS and at high redshifts.

The accuracy of the $Ar_3O_3$,  $S_3O_3$ and $N_2$ metallicity indicators is comparable at low metallicity, and the former two are more accurate at high metallicity, as can be judged  from the distribution of points between the thin curves in Figs. 1a, 1b and 2a. But, as mentioned above, for \emph {integrated galaxy spectra},  $Ar_3O_3$ and  $S_3O_3$  should be superior even at low metallicity (at least they do not call for a different calibration  to take into account the diffuse ionized medium). On the other hand, both $Ar_3O_3$  and $S_3O_3$ demand a reliable reddening correction, which is not required for $N_2$. 
  In the case of global spectra of galaxies, reddening correction can be done empirically  by using  the  observed relation  between  the Balmer extinction and the galaxy excitation class or the discontinuity at 4000\AA\ (Stasi\'nska et al. 2004).  This avoids the measurements of \Ha\ and \Hb\ l intensities, often somewhat problematic.. 

For the high metallicity end, our calibrations of $Ar_3O_3$,  $S_3O_3$ rely on the $P$-method of Pilyugin (2001a).  When metal-rich HII regions become better understood, a recalibration of all the metallicity indexes in a consistent way may be needed. However, our discussion of the merits and drawbacks of  $Ar_3O_3$,  $S_3O_3$ with respect to other methods will remain valid.

\begin{acknowledgements}
Many thanks to Yuri Izotov, Fabio Bresolin, Leonid Pilyugin and the referee, Jose Vilchez, for useful comments.

\end{acknowledgements}

\end{document}